\documentclass[12pt]{preprint}

\usepackage{psfig}
\usepackage{amssymb} 
\usepackage{symbols}

\newcommand{\vers}{25.10.95}

\newcommand{\Real}{\mbox{l} \! \mbox{R}}
\newcommand{\cbc}{\chb\chi}

\begin{document}
\dateandnumber(October 1995){J{\"u}lich,
  HLRZ 46/95}{hep-lat/9511005}%
\titleofpreprint%
{             Interplay of universality classes                }%
{            in a three-dimensional Yukawa model$^*$           }%
{                                                              }%
{                                                              }%
{                                                              }%
\listofauthors%
{  E.~Focht$^{1,2}$, J.~Jers{\'a}k$^{1,3}$ and J.~Paul$^{1,4}$ }%
{                                                              }%
{                                                              }%
\listofaddresses%
{\em $^1$Institute of Theoretical Physics E,
  RWTH Aachen,         D-52056 Aachen, Germany                 }%
{\em   and~ HLRZ c/o KFA J{\"u}lich,
      D-52425 J{\"u}lich, Germany                              }%
{
                                                               }%
{}

\abstractofpreprint{
We investigate numerically on the lattice the interplay of
universality classes of the three-dimensional Yukawa model with U(1)
chiral symmetry, using the Binder method of finite size scaling. At
zero Yukawa coupling the scaling related to the magnetic Wilson--Fisher
fixed point is  confirmed. At sufficiently strong Yukawa coupling the
dominance of the chiral fixed point associated with the 3D Gross--Neveu model
is observed for various values of the coupling parameters, including
infinite scalar selfcoupling. In both cases the Binder method works
consistently in a broad range of lattice sizes. However, when the
Yukawa coupling is decreased the finite size behavior gets complicated
and the Binder method gives inconsistent
results for different lattice sizes. This signals a cross-over between
the universality classes of the two fixed points.
%
}

\footnoteoftitle{
\footnoteitem($^*$){ \sloppy
Supported
by BMBF and DFG.
}
\footnoteitem($^2$){ \sloppy
E-mail address: focht@hlrz.kfa-juelich.de
}
\footnoteitem($^3$){ \sloppy
E-mail address: jersak@physik.rwth-aachen.de
}
\footnoteitem($^4$){ \sloppy
E-mail address: jpaul@hlrz.kfa-juelich.de
}
}
%
\pagebreak
%

\section{Introduction}

Some strongly coupled lattice field theories in 4 dimensions (4D)
possess perturbatively unaccessible critical points where scaling
properties are understood only poorly or not at all.  Examples are
noncompact QED \cite{recREFS1}, compact QED without matter fields
(pure QED) \cite{JeLa95} or with fermions \cite{Ok89}, gauged
Nambu--Jona-Lasinio or Yukawa models \cite{recREFS2} and models with
fermions, gauge field and charged scalar at strong gauge coupling
\cite{recREFS3}.  A clarification of their critical behavior and of
the continuum limit taken at such points is desirable at least for two
reasons: Firstly, the fundamental question of the existence of 4D
quantum field theories defined on nongaussian fixed points has never
been settled.  Secondly, finding a 4D theory interacting strongly at
short distances could contribute to the development of theoretical
scenarios for dynamical symmetry breaking as possible alternatives to
the Higgs mechanism in the standard model and its extensions.

Except the pure QED, a chiral phase transition, with the chiral
condensate $\lag\cbc\rag$ as an order parameter, takes place at all
the critical points mentioned above.  But it always differs in some
qualitative way from the classical model for chiral symmetry breaking,
the Nambu--Jona-Lasinio model.  This is encouraging, as that model is
even nonperturbatively nonrenormalizable \cite{AlGo95} and thus of
very limited use.  The differences consist mainly in an admixture of
some other phenomena like confinement, monopoles, magnetic or Higgs
transition, additional states of vanishing mass, etc., intertwining
with the chiral transition, but occuring also in other situations,
including those without fermions.  This increases the hope for a
fundamental difference from the Nambu--Jona-Lasinio model, but also
makes the transitions perplexingly complex and difficult to analyze.
In particular, the genuine character of the transition might be hidden
behind some prescaling phenomena caused by some component of the
mixture, or by a crossover between different universality classes.

In this paper we study the interplay of the chiral and magnetic phase
transitions in a 3D lattice Yukawa model (Y$_3$ model) with global
U(1) chiral symmetry as an exercise for the investigations of
analogous but more complex situations in 4D.  The Y$_3$ model has
nontrivial fixed points, a property searched for in 4D.  We would like
to learn how to detect such points, and what are the possible
obstacles when the scaling properties are investigated numerically in
the situation of intertwining phenomena.

\begin{figure}
  \begin{center}
    \leavevmode
    \psfig{file=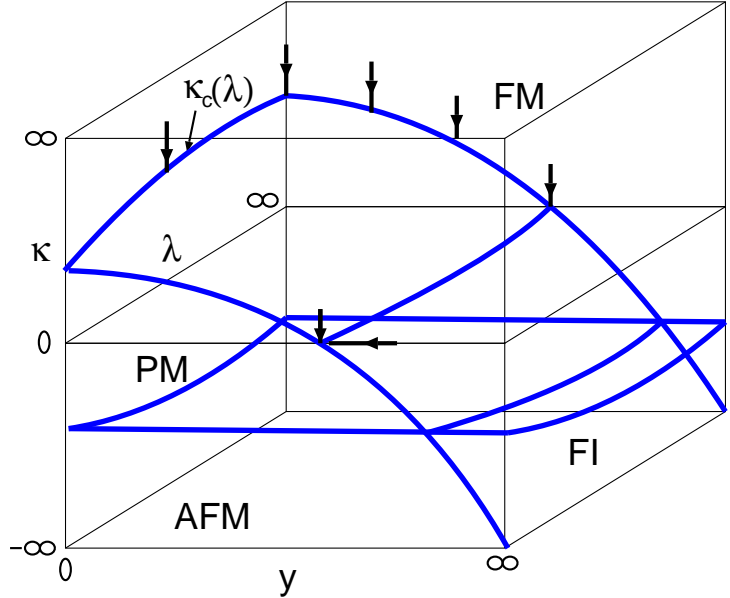,height=8cm}
  \end{center}
  \caption{ Schematic phase diagram of the Y$_3$ model. The region
    below the upper critical surface is the paramagnetic phase (PM),
    the region above it the ferromagnetic phase (FM). The $y=0$ plane
    and the $\kp=\lm=0$ line correspond to the $\phi_3^4$ and GN$_3$
    models, respectively. For negative values of the parameter
    $\kappa$ we further expect an antiferromagnetic phase (AFM) and a
    ferrimagnetic phase (FI). We have investigated the PM-FM
    transition for $\kp\simeq 0$ and $\kp>0$, in particular at the
    points and directions indicated by the arrows.}
  \label{Phasediag}
\end{figure}

\begin{figure}[htbp]
  \begin{center}
    \leavevmode
    \psfig{file=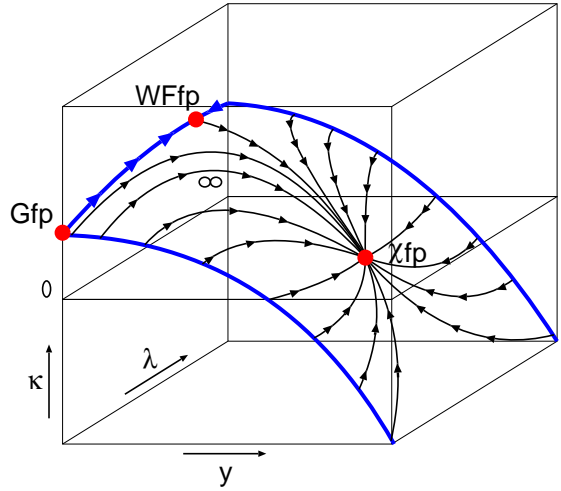,angle=270,height=6.5cm}
  \end{center}
  \caption{A suggestion for the renormalization group flow on the
    PM-FM critical surface of the Y$_3$ model. The fixed points are
    Gfp (Gaussian), WFfp (Wilson-Fisher) and $\chi$fp (chiral, or
    GN$_3$). The indicated position of the $\chi$fp is very schematic,
    it could lie anywhere on the PM-FM sheet, at $y>0$.}
  \label{FIG:RGFL}
\end{figure}

The couplings of the Y$_3$ model are the scalar hopping parameter
$\kp$, the scalar quartic selfcoupling $\lm$ and the Yukawa coupling
$y$. The action is given in subsec.~2.1. The phase diagram is shown
schematically in fig.~\ref{Phasediag}. We concentrate on the
transition between the paramagnetic (PM) and ferromagnetic (FM)
phases. The two-dimensional PM-FM sheet of 2$^{\rm nd}$ order phase
transitions connects the critical line of the purely scalar $\phi_3^4$
model at $y=0$ and the critical point of the 3D Gross-Neveu (GN$_3$)
model at $\kp=\lm=0$. On this sheet the Y$_3$ model is expected to
have two nontrivial fixed points:
\begin{enumerate}
\item Wilson-Fisher fixed point (WFfp) \cite{WiFi72} of the pure
  scalar 2-component $\phi_3^4$ theory, whose most familiar
  representative is the 3D XY (XY$_3$) model.  The phase transition is
  of magnetic type.
\item Chiral fixed point ($\chi$fp), most naturally associated with
  the GN$_3$ model with U(1) global chiral symmetry and a chiral
  phase transition.  The existence of this fixed point is related to
  the nonperturbative renormalizability of the GN$_3$ model
  (see \cite{RoWa91} and references therein).
\end{enumerate}
The sketch of the renormalization group flow in fig.~\ref{FIG:RGFL}
represents a plausible scenario for what happens along the critical
PM-FM sheet: The magnetic WFfp describes only the $\phi_3^4$ theory.
The $\chi$fp presumably dominates (has a domain of attractivity)
everywhere as long as the Yukawa coupling does not vanish, and in the
limit of infinite cutoff the Y$_3$ model is thus equivalent to the
GN$_3$ model. This expectation has been recently supported at weak
scalar selfcoupling and large Yukawa coupling by the 1/N expansion
\cite{Zi91,KoRo91,GaKo92} and a consequent combined analytic and
numerical investigation \cite{KaLa93}. A discussion of the equivalence
between the Yukawa and four-fermion theories, as well as earlier
references, can be found in ref.~\cite{HaKo93}.

\begin{figure}[htbp]
  \begin{center}
    \leavevmode
    \psfig{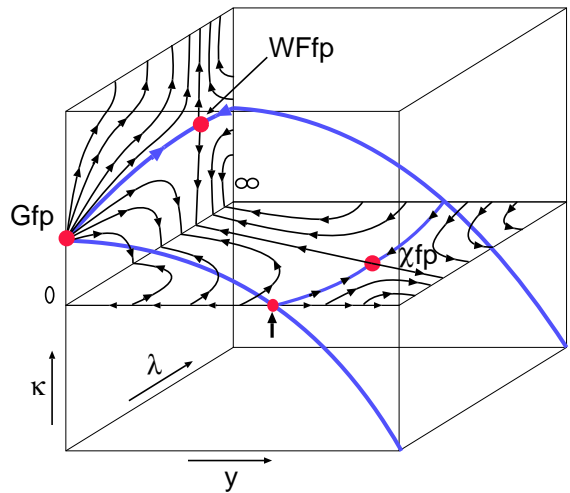}
  \end{center}
  \caption{The schematic RG-flow in the $\phi_3^4$ model ($y=0$, $\kp>0$),
    in the $\kp=0$ surface of the Y$_3$ model and on the $\kp=\lm=0$
    line, which corresponds to the GN$_3$ model. The fixed point of
    the latter model is indicated by an arrow.}
  \label{FIG:RGFL2}
\end{figure}

In fig.~\ref{FIG:RGFL2} we show schematic RG flows also
outside the critical sheet for three special cases of restricted
parameter space: $y=0$, $\kp=0$ and $\kp=\lm=0$. This figure indicates
that the known RG flows in the $\phi_3^4$ and GN$_3$ models can be
consistently embedded into the RG flows in the Y$_3$ model.

When in the Y$_3$ model the Yukawa coupling decreases and the
$\phi^4_3$ theory is approached, the WFfp gets presumably influential,
as some crossover to the magnetic universality class must occur.  This
consideration warns us that for limited lattice volumina and
consequently limited correlation lengths either no unique finite size
scaling behavior can be found or the wrong fixed point dominates.
Thus a detection of the genuine -- presumably chiral -- character of
the transition gets more and more difficult in numerical simulations.
This is the situation we are most interested in, as it might occur in
4D without a prior warning.

Apart from this particle physics motivation our work might be of interest
also for other reasons, related mainly to statistical mechanics:
\begin{enumerate}
\item We have applied the Binder method of finite size scaling
  analysis \cite{Bi81,Bi92} to the chiral phase transition and found
  that it works very well also when a composite scalar field is used
  in the finite size scaling theory, as long as the $\chi$fp alone
  dominates the finite size scaling behavior.
\item A transition between various universality classes in finite
  volumina has been investigated recently \cite{BiDe92,DORe94} in some
  spin models, but, to our knowledge, until now in no models with
  fermions. Thus we make a new contribution to the experience with
  this sort of complex finite size behavior. As in spin models, it is
  the failure of the Binder method which indicates a change of the
  universality class.
\item Sometimes an intermediate universality class could exist
  \cite{DORe94}.  This would be very surprising for the Y$_3$ model,
  nevertheless we have verified that this is most probably not the
  case here.
\end{enumerate}

We now briefly describe the contents of the paper and the main results:

In the next section we introduce the Y$_3$ model
and determine its phase diagram (fig. \ref{Phasediag}), both by means
of the effective potential
in the one loop approximation, and by performing numerical simulations
on a small lattice at many points in the three-dimensional
parameter space.
The most useful order parameter is the scalar field expectation value,
even if this field can be considered as composed of a fermion pair.
We mention some results on the fermion and boson masses
both in the symmetric phase and in the phase with broken chiral symmetry.

In sec.~3 we shortly review the Binder method allowing a determination
of several critical exponents by an analysis of finite size effects.
The most useful exponent is the correlation length exponent $\nu$
obtained from the Binder-Challa-Landau (BCL) \cite{Bi81,BiCh86} cumulant.

The magnetic transition of the $\phi^4_3$ theory is investigated in
sec.~4.  After localizing the critical line we concentrate on the case
$\lm=\infty$ (the XY$_3$ model) and a case of an intermediate scalar
selfcoupling ($\lm=0.5$).  The obtained exponents are consistent with each
other and with the value expected from analytic investigations of the WFfp
($\nu=0.67$).  Also the values of the renormalized coupling extrapolated to
infinite cutoff are consistent.  The Binder method is compared with two other
approaches to finite size scaling and found to be most suitable for our
purposes.

Sec.~5 deals with the chiral transition in the GN$_3$ model
at $\lm=0$ both in the auxiliary scalar field formulation
($\kp = 0$), and with a dynamical scalar field ($\kp$ varied
and $y$ kept at the critical value, $y=1.09$).
In both approaches to the critical point the Binder method works
comparably well for all the lattice sizes we used (6$^3$ -- 24$^3$)
and gives consistent results for critical exponents.
In particular, $\nu = 1.03(11)$, which is
a value consistent with theoretical expectations \cite{HaKo93,HeKu92,Gr94}
and significantly different from the value found
for the $\phi^4_3$ model at $y = 0$.
Thus the difference between the magnetic and chiral universality
classes is clearly observed in the $y=0$ and $\lm=0$ limit cases.
Their common property is that the Binder method works in an
exemplary way in the whole range of lattice sizes we used.

In sec.~6, the Y$_3$ model with a large Yukawa coupling, $y = 1.1$, is
investigated at the maximal value of the scalar selfcoupling $\lm =
\infty$.  Also here the Binder method works quite well, and we find
$\nu = 0.88(6)$, a value slightly lower than, but within errors still
consistent with that found in the GN$_3$ model.  This confirms the
appurtenance of the Y$_3$ model with both couplings $y$ and $\lm$
strong to the same chiral universality class as the GN$_3$ model, and
thus the physical equivalence of both theories.

However, difficulties arise when the Yukawa coupling decreases.  As we
describe in sec.~7, at $\lm = \infty$ and $y = 0.6$ the BCL cumulants
cross at different points when only small (6$^3$ -- 10$^3$) or large
(10$^3$ -- 24$^3$) lattices are considered, suggesting different
values of the critical $\kp$.  Restricting ourselves to the larger
lattices only, we find the Binder method to work, giving $\nu =
0.99(23)$.  This value is consistent with the GN$_3$ model value, but
has a large error.  On small lattices the obtained value of $\nu$ is
significantly lower and close to the value in the $\phi^4_3$ model.
As we describe in detail in the same section, at $\lm = \infty$ and $y
= 0.3$ the Binder method gives inconsistent results in the whole range
of lattice sizes 6$^3$ -- 32$^3$ we have investigated.  This can be
interpreted as a situation in which none of the two fixed points alone
dominates the finite size effects on lattices of these sizes, i.e. as
an interplay of or a crossover between universality classes.  We find
no sign for the existence of an intermediate universality class.

As we conclude in sec.~8, an interplay of magnetic and chiral
phenomena in the Y$_3$ model thus results in uncontrollable finite
size effects.  However, inconsistencies in the application of finite
size methods become apparent only when a broader range of lattice
sizes is investigated.  This might serve as a warning for
investigations of critical points with a mixture of chiral and some
other critical behaviour in 4D lattice field theories.

\section{The $\mbox{Y}_3$ model and its phase diagram}

\subsection{The action}

In order to investigate
the breakdown of a continuous chiral symmetry we use
staggered fermions \cite{Su77}. In the lattice parametrization the
action of the Y$_3$ model is
\begin{eqnarray} \label{Gitterwirk}
S &=& S_{\mathrm{B}} + S_{\mathrm{F}} + S_{\mathrm{Y}} \nonumber \\
S_{\mathrm{B}} &=& \sum_{x} \left\{ -2 \kappa \sum\limits_{\mu}
\sum\limits_{i=1}^{2}
\phi_{x+ \mu}^{i} \phi_x^{i} + \sum\limits_{i=1}^{2}
\left(\phi_x^{i} \right)^2 + \lambda \left( \sum\limits_{i=1}^{2}
(\phi_x^{i})^2 - 1 \right)^2 \right\} \nonumber \\
S_{\mathrm{F}} &=& \frac{1}{2} \sum\limits_{x,\mu} \eta_{x,\mu}
\sum\limits_{j=1}^{N_F/2} \left(
\bar{\chi}_x^{j} \chi_{x + \mu}^{j} - \bar{\chi}_{x + \mu}^{j}
\chi_x^{j} \right) \nonumber \\
S_{\mathrm{Y}} &=& \frac{y}{2^3} \sum\limits_{x,j} \bar{\chi}_x^{j}
\sum\limits_b \left(
  \phi_{x+b}^1 + i \varepsilon_x \phi_{x+b}^2 \right) \chi_x^{j} \: ,
\end{eqnarray}
where the integer 3-vectors $x$, $x+\mu$ and $x+b$ denote, respectively,
lattice site, its nearest neighbors and corners of the associated elementary
cube (both in positive direction). The coefficients are:
\begin{displaymath}
\eta_{x,1} = 1 \: , \;
\eta_{x,\mu} = (-1)^{x_1 + \ldots + x_{\mu - 1}} \: ,\;
\varepsilon_x = (-1)^{x_1 + \ldots + x_3} \; .
\end{displaymath}
The coupling constants $\kappa$, $\lambda$ and $y$ and
the fields $\phi^{i}$ and $\chi_j$ are dimensionless
quantities. $N_F=4$ is the number of continuum four-component fermions.

The scalar sector $S_B$ of the action (\ref{Gitterwirk}) has a global
O(2) symmetry.  The action $S$ is invariant under the vectorial U(N)
transformations
\begin{equation}
\chi_{j} \rightarrow \Omega_{ji} \chi_{i} \,\, , \,\, \bar{\chi}_j
\rightarrow \bar{\chi}_i \Omega_{ij}^{\dagger} \,\, , \,\, \Omega \in
U(N_F/2)
\end{equation}
and the axial $\mbox{U}(1)_{\mathrm{A}}$ transformations
\begin{equation}
 \chi \rightarrow e^{i \omega_{\mathrm{A}} \varepsilon_x} \chi
 \,\, , \,\, \bar{\chi}
 \rightarrow \bar{\chi} e^{i \omega_{\mathrm{A}} \varepsilon_x} \, \, ,
 \,\, \phi \rightarrow
 e^{-2i \omega_{\mathrm{A}}} \phi \,\, , \, \,\phi^* \rightarrow
 e^{2i \omega_{\mathrm{A}}} \phi^* \; , \; \omega_{\mathrm{A}} \in \Real
 \; .
\end{equation}

The action (\ref{Gitterwirk}) contains two important limit cases, the
$\phi_3^4$ model and the GN$_3$ model. At $y=0$ it is the $\phi_3^4$
theory described by the purely scalar part $S_{\mathrm{B}}$ of
(\ref{Gitterwirk}). In the limit $\lambda \rightarrow \infty$ the
action $S_B$ reduces to that of the XY$_3$ spin model. At
$\kappa = \lambda = 0$, the action $S$ (\ref{Gitterwirk}) turns into
the action of the chiral GN$_3$ model in the auxiliary scalar field
formulation. The full Yukawa model interpolates between both these
models and the PM-FM critical sheet continuously connects the magnetic
phase transition of the spin model with the chiral phase transition of
the GN$_3$ model.

\subsection{Symmetry breaking}

In order to get information about the breakdown of the
continuous chiral symmetry in the $\mathrm{Y}_3$ model we have computed
the effective potential in 1-loop order for $\kp>0$.
For this purpose we start with the Euclidean continuum
action with $m_0$ being the bare mass, $g_0$ the bare scalar
selfcoupling and $y_0$ the bare Yukawa coupling. The calculation is
straightforward (see \cite{ItZu80}) and yields
\begin{eqnarray} \label{EffPot}
V_{\mathrm{eff}}(\sigma^2) &=& \frac{m_0^2}{2} \sigma^2 +
\frac{g_0}{4!} (\sigma^2)^2
+  \frac{1}{2} \int\limits_{\Lambda} \frac{d^3p}{(2\pi)^3} \,
\ln \left( p^2 + m_0^2 + \frac{g_0}{2} \sigma^2 \right)  \nonumber \\
 & + & \frac{1}{2} \int\limits_{\Lambda} \frac{d^3p}{(2\pi)^3} \,
\ln \left( p^2 + m_0^2 + \frac{g_0}{6} \sigma^2 \right) \nonumber \\
&-& 2 N_F \int\limits_{\Lambda} \frac{d^3p}{(2\pi)^3} \ln
  \left( p^2 + y_0^2 \sigma^2 \right) \; ,
\end{eqnarray}
where we have regularized the momentum integrals with a
cut-off $\Lambda$.
We have introduced the abbreviation $\sigma^2 = \sigma_1^2 +
\sigma_2^2$, where the constants $\sigma_i$ ($i=1,2$) can be identified
with the expectation values $\sigma_i = \langle \varphi_i \rangle$,
$\varphi_i$ being the scalar fields in the continuum. These fields are
related to the lattice scalar fields $\phi_x^i$ by
\begin{equation}
\varphi_i(ax)=\sqrt{\frac{2\kp}{a}}\phi_x^i
\end{equation}
and the relations between the parameters are
\begin{equation} \label{EQ:C2L}
  m_0^2=\frac{1-2\lm-6\kp}{a^2\kp}~,~~
  g_0=\frac{6\lm}{a\kp^2}~.
\end{equation}

All the values of $\sigma_i$ which minimize
$V_{\mathrm{eff}}$ are possible candidates for the vacuum of the
theory. We can find these minima by solving the equations $\partial
V_{\mathrm{eff}} /\partial \sigma_i = 0$ simultaneously for $ i=1,2$.
One solution is $\sigma_1 = \sigma_2 = 0$. In the symmetric phase it
is a minimum  ($\partial_i \partial_i V_{\mathrm{eff}}|_{(0,0)} > 0$),
in the broken phase a maximum ($\partial_i \partial_i
V_{\mathrm{eff}}|_{(0,0)} < 0$) and a further solution exists.
In this sense $\partial_i \partial_i V_{\mathrm{eff}}|_{(0,0)} = 0$ is
an implicit equation for the boundary between both phases of the theory.

At fixed values of the parameters $m_0^2$ and $g_0$ we can
calculate the critical Yukawa coupling $y_{\mathrm{c}}(m_0^2, g_0)$. If
we choose $m_0^2 \ge 0$ we can always find a positive solution $y_c$ of
$\partial_i \partial_i V_{\mathrm{eff}}|_{(0,0)} = 0$ which is
\begin{equation} \label{criticalYukawa}
y_{\mathrm{c}} = \left(4 N_F \int\limits_\Lambda \frac{d^3p}{(2\pi)^3} \,
  \frac{1}{p^2} \right)^{-\frac{1}{2}} \left[ m_0^2 + \frac{2}{3}
\int\limits_\Lambda \frac{d^3p}{(2\pi)^3} \, \frac{g_0}{p^2 + m_0^2}
\right]^{\frac{1}{2}} \; , \; g_0 \ge 0 \; .
\end{equation}
Equation (\ref{criticalYukawa}) means that even for $m_0^2\geq 0$,
when the classical potential does not predict the symmetry breaking, a
solution $y_{\mathrm{c}}(m_0^2, g_0)$ exists. For all couplings
$y_0$ with $y_0 > y_{\mathrm{c}}$ the vacuum expectation value
$\langle \varphi \rangle$ of the scalar field is nonzero and the
chiral symmetry is broken.

This computation of the 1-loop effective potential suggests
that in the $\mathrm{Y}_3$ model, at sufficiently small $\kp\geq 0$, two
phases of different symmetry exist, as indicated in
fig.~\ref{Phasediag}. As usual, we call them paramagnetic (PM) for
$\lag\phi\rag=0$ and ferromagnetic (FM) for $\lag\phi\rag\ne 0$.

\subsection{The phase diagram}

Figure \ref{Phasediag} displays a schematic phase diagram, including
also some expectations for $\kp<0$.  The phases relevant for our
purposes are PM and FM. In the PM phase both order
parameters $\langle \phi \rangle$ and $\langle \bar{\chi} \chi
\rangle$ (for $y>0$) are zero and fermions are massless. The
lightest boson pair is degenerate.  In the FM phase the vacuum
expectation value of the scalar field, the chiral condensate and
fermion mass are nonzero.

To characterize the PM and FM phases numerically we have used
the magnetization $M = V^{-1} \sqrt{ (\sum_x \phi_x^1 )^2 + (\sum_x
  \phi_x^2 )^2 }$, $V$ being the number of
lattice points.
A continuous phase transition is indicated by a singularity of the
susceptibility
\begin{equation}
\chi = V ( \langle M^2 \rangle - \langle M \rangle^2 ) \; .
\end{equation}

For the numerical simulations we used the Hybrid Monte Carlo
algorithm. The critical surface of the phase diagram has been found
by localizing peaks of the susceptibility on a $6^3$-lattice.
In table \ref{Peaks} the values of the coupling parameters at the
maxima of the susceptibility are summarized. Of course, they give
only an approximate position of the critical surface.  In cases in which it
was needed, the critical coupling in the thermodynamic limit has been
determined by a finite size scaling analysis.

\begin{table}[htb]
  \caption{ Peaks of the susceptibility in the
    Y$_3$ model determined on a $6^3$ lattice}
  \label{Peaks}
  \begin{center}
    \parbox{7cm}{
      \begin{tabular}{||c|c|c||} \hline\hline
$y$ & $\kappa$ & $\lambda$ \\ \hline\hline
0 & 1/6 & 0 \\ \hline
0 & 0.1730(6) & 0.01 \\ \hline
0 & 0.1818(5) & 0.03 \\ \hline
0 & 0.2008(10) & 0.10 \\ \hline
0 & 0.2265(15) & 0.30 \\ \hline
0 & 0.238(2) & 0.50 \\ \hline
0 & 0.249(2) & 1.00 \\ \hline
0 & 0.240(3) & 3.00 \\ \hline
0 & 0.218(2) & $\infty$ \\ \hline
1.10(8) & 0 & 0.0 \\ \hline
1.25(8) & 0 & 0.5 \\ \hline
1.25(10) & 0 & 0.75 \\ \hline
1.28(8) & 0 & 1.0 \\ \hline
1.25(13) & 0 & 1.5 \\ \hline
1.25((13) & 0 & 2.0 \\ \hline
1.10(8) & 0 & $\infty$ \\ \hline
0.3 & 0.15(1) & 0 \\ \hline
0.6 & 0.12(1) & 0 \\ \hline
0.80(8) & 0.08 & 0 \\ \hline
\end{tabular}
}
\parbox{7cm}{
\begin{tabular}{||c|c|c||} \hline\hline
$y$ & $\kappa$ & $\lambda$ \\ \hline\hline
0.90(8) & 0.05 & 0 \\ \hline
1.42(7) & -0.1 & 0 \\ \hline
0.0 & 0.24(2) & 0.5 \\ \hline
0.3 & 0.22(1) & 0.5 \\ \hline
0.5 & 0.20(2) & 0.5 \\ \hline
0.95(13) & 0.1 & 0.5 \\ \hline
1.10(10) & 0.6 & 0.5 \\ \hline
1.30(13) & 0.0 & 0.5 \\ \hline
0.0 & 0.25(4) & 1.0 \\ \hline
0.3 & 0.22(3) & 1.0 \\ \hline
1.05(15) & 0.06 & 1.0 \\ \hline
1.30(25) & 0.0 & 1.0 \\ \hline
0.0 & 0.22(1) & $\infty$ \\ \hline
0.3 & 0.19(1) & $\infty$ \\ \hline
0.6 & 0.14(1) & $\infty$ \\ \hline
0.80(8) & 0.1 & $\infty$ \\ \hline
1.0 & 0.04(1) & $\infty$ \\ \hline
1.10(8) & 0.001 & $\infty$ \\ \hline
1.47(8) & -0.1 & $\infty$ \\ \hline
\end{tabular}
}
\end{center}
\end{table}

Our data strongly supports the expectation that for all positive
values of $y$ the condensate $\langle \bar{\chi} \chi \rangle$
vanishes simultaneously with the magnetisation $M$.  We have extracted
the fermion mass from the fermionic momentum space propagator. The
agreement with the tree level prediction $am_F=y\lag\phi\rag$
is quite good.  In the FM phase we have also observed in the
$\phi$-propagator a massive
particle, the $\sigma$-boson, and a massless particle, the Goldstone
boson. These masses, as well as $\lag\bar\chi\chi\rag$, are not as
convenient as the BCL cumulant for the study of the finite size
behavior but can be used for a qualitative comparison of the physical
content of the Y$_3$ model in different parameter regions.

\subsection{Renormalizability properties}

Both the $\phi^4$ theory and the full Yukawa model in 3D are perturbatively
superrenormalizable.  For the GN$_3$ model this is different. The continuum
4-fermion coupling has negative mass dimension, and the corresponding
interaction is therefore perturbatively nonrenormalizable. Nevertheless, it has
been proved that the GN$_3$ model is renormalizable in the $1/N_F$-expansion
\cite{PaRo91}.

It has also been shown in the framework of $1/N_F$-expansion that for
weak scalar selfcoupling $\lambda = O(1/N_F)$ the Gross-Neveu model
and the full Yukawa model in $2 < d < 4$ are equivalent field theories
\cite{Zi91,KoRo91,GaKo92}. Near the nontrivial fixed point the kinetic
term of the scalar field and the quartic scalar selfinteraction turn
out to be irrelevant operators.  However, in those works nothing
beyond the range of validity of the $1/N_F$-expansion could be said.

In ref.~\cite{KaLa93} the equivalence has been confirmed by analytic and
numerical methods for the discrete chiral Z(2)-symmetry, still with
$\lm=O(1/N_F)$. We have extended that work to the U(1)-symmetric case
and have investigated a wide range of parameters including infinite
scalar selfcoupling.

\section{Finite size scaling theory}

\subsection{The Binder method}

In order to examine the interplay of the universality classes associated with
two different nontrivial fixed points in the Y$_3$ model we have studied the
finite size scaling behavior and tried to determine the critical exponents of
the theory\footnote{The critical exponents $\nu$, $\beta$ and $\gamma$ are
  defined as follows ($t$ is the reduced coupling):
\begin{displaymath}
  \xi\sim |t|^{-\nu},~ M\sim t^\beta~(\mbox{for } t>0),
  ~\chi\sim |t|^{-\gamma}.
\end{displaymath}}
at several points of the critical surface.  A very powerful method to
do this is the Binder method of finite size scaling analysis of a
cumulant \cite{Bi81,Bi92}.

It is sufficient to use scalar n-point
functions even in the case of nonvanishing Yukawa coupling.
We therefore follow refs.~\cite{Bi92,BeGo88a} and define the corresponding
fourth-order cumulant $U_L$ on a cubic lattice of extent $L$:
\begin{equation} \label{BiKuON}
U_L = -\frac{\frac{1}{V} \tilde{G}_L^{(4)} - 2
  \left[\tilde{G}_L^{(2)} \right]^2}{\left[ \tilde{G}_L^{(2)}
  \right]^2} \; , \; V = L^3 \; ,
\end{equation}
where $\tilde{G}_L^{(2)}$ and $\tilde{G}_L^{(4)}$ are
\begin{eqnarray}
\tilde{G}_L^{(2)} &=& \frac{1}{V} \sum\limits_{x_1,x_2}
\sum\limits_{i} \langle
\phi_{x_1}^{i} \phi_{x_2}^{i} \rangle \nonumber \\
\tilde{G}_L^{(4)} &=& \frac{1}{V} \sum\limits_{x_1,\ldots,x_4}
\sum\limits_{i,j} \langle \phi_{x_1}^{i} \phi_{x_2}^{i} \phi_{x_3}^{j}
\phi_{x_4}^{j} \rangle \; .
\end{eqnarray}
If both $L$ and the correlation length $\xi$ are sufficiently large
then $U_L$ has the form
\begin{equation} \label{BiKuU1b}
U_L = 2 - \frac{f_4 \left( \frac{L}{\xi} \right)}{\left[ f_2 \left(
    \frac{L}{\xi} \right) \right]^2 } \;
\end{equation}
with analytic functions $f_2$ and $f_4$. Note that (\ref{BiKuU1b})
requires the validity of hyperscaling.

At the critical value $\kp_c$ of the hopping parameter $\kp$ the correlation
length diverges and all cumulants $U_L|_{\kappa_{\mathrm{c}}}$ have the
same value $U^*$ independent of the lattice size. This makes it possible to
determine the infinite volume critical coupling as the common intersection
point of $U_L$ for different values of $L$.

In the scaling limit Binder's cumulant has the form
\begin{equation} \label{Kumulhelp}
U_L(t) = U \left( \frac{L}{\xi(t)} \right) = \tilde{U} ( t L^\frac{1}{\nu} )
\;\; , \; \; |t| \ll 1
\end{equation}
with $t= 1 -\kappa_\mathrm{c}/\kappa$.
Let us consider a pair $(bL,L)$ of lattice sizes with $b > 1$. From
(\ref{Kumulhelp}) it follows
\begin{equation} \label{NuBinderA}
\ln \left( \frac{\partial U_{bL}}{\partial
    U_L}\bigg|_{\kappa_{\mathrm{c}}} \right) = \frac{1}{\nu} \ln b\; .
\end{equation}
In order to obtain the derivative $\frac{\partial
  U_{bL}}{\partial U_L}|_{\kappa_\mathrm{c}}$ one calculates the
function $U_{bL} = g(U_L)$
numerically and near the critical point approximates $g$ by a
linear function determining its slope.

Similar relations can easily be derived for the exponent $\gamma$ of
the susceptibility $\chi$ and the exponent $\beta$ of the
magnetisation $M$,
\begin{eqnarray} \label{BinderScal}
\ln \left( \frac{\chi_{bL}( \kappa_{\mathrm{c}})}{\chi_L (
    \kappa_{\mathrm{c}})} \right) &=& \frac{\gamma}{\nu} \ln b \nonumber \\
\ln \left( \frac{M_{bL}( \kappa_{\mathrm{c}})}{M_L (
    \kappa_{\mathrm{c}})} \right) &=& -\frac{\beta}{\nu} \ln b  \; .
\end{eqnarray}
Using (\ref{BinderScal}) one can calculate the ratios $\beta/\nu$ and
$\gamma/\nu$ from $M$ and $\chi$ determined on various lattice sizes
$(bL,L)$ exactly at $\kappa_{\mathrm{c}}$.

In the $\phi_3^4$ theory the specific heat exponent $\alpha$ is negative. That
means that the specific heat is a regular function of the reduced coupling and
there is no relation similar to (\ref{BinderScal}) for it.

To calculate the required quantities, we used a reweighting technique.
By means of the original method due to Ferrenberg and Swendsen \cite{FeSw89}
one can only interpolate operators which can be expressed as explicite
functions of $S$. Therefore, like the authors of ref.~\cite{KaLa93}, we used a
variation of the method suggested in ref.~\cite{KaKa91}. It can be regarded as
the multihistogram method with bins of zero width. With that reweighting
technique one can interpolate nearly arbitrary operators over a
wide range of the coupling $\beta$. For this purpose it is necessary to store
the operator $S$ which corresponds to the coupling $\beta$ and the value of
the operator for each configuration which has been generated during
the simulation.

\subsection{Previous applications of the Binder method}

In the past the Binder method has been applied to a variety of
interesting physical systems.  In ref. \cite{BeGo88a} the method has
been generalized to $O(N)$ $\phi^4$ theories and in ref.
\cite{BeGo88b} the critical exponent $\nu$ has been determined for the
$\mathrm{O}(4)$ invariant scalar $\phi^4$ theory in 3D and 4D. A high
precision measurement of $\nu$ in the XY$_3$ model has been done
in ref.~\cite{Ja90}.

The method has also been applied to models with interacting fermions
\cite{KaLa93}. Here a slight modification of the Binder method has
been used to compute the critical exponents $\nu$ and $\gamma /\nu$ in
the 3-dimensional Gross-Neveu model with Z(2)-symmetry. The found
value of $\nu \approx 1$ is in good agreement with the prediction of
the $1/N_F$-expansion.

In ref. \cite{DORe94} the critical behaviour of diluted Heisenberg
ferromagnets with competing interactions has been investigated. The
authors varied the concentration of spins and found two distinct
universality classes which are separated by a crossover region. In
this domain strong corrections to scaling appear, and Binder's method
does not work well. Also evidence for a new, intermediate universality
class has been found.

\subsection{Other methods to determine critical exponents}

For the $\phi_3^4$ model we have also tried to compute critical
indices by some other methods. Among these are the direct method,
which makes use of the finite size scaling laws of physical
quantities, and the scaling of the smallest Lee-Yang zero with the
lattice size.

On a finite lattice of extent $L$ the susceptibility peaks at the
value $\kappa_M(L)$ of the hopping parameter. If we increase the
lattice size then $\kp_M(L)$ approaches $\kp_c$ according to
\begin{equation}
\kp_M(L)-\kp_c ~ \propto ~ L^{-1/\nu}.
\end{equation}
Thus the measurement of $\kp_M(L)$ for various lattice sizes $L$
yields the critical exponent $\nu$ by a corresponding fit. We have
tried this method in the $\phi_3^4$ theory for different values of the
scalar selfcoupling $\lambda$. Our results were rather unsatisfactory
because of their quite large statistical errors. For the same
statistics we obtained more accurate values for $\nu$ with the Binder
method.

Another possibility to determine $\nu$ is to use the finite size scaling
of the Lee-Yang-Fisher zeroes. By continuing the
hopping parameter $\kappa$ to complex values one finds that the
partition function has zeroes in the complex plain. For finite
lattices all the zeroes lie off the real axis.
The zero $\kappa_0$ with the smallest
imaginary part scales like \cite{ItPe83}
\begin{displaymath}
\mathrm{Im}(\kappa_0) \sim L^{-\frac{1}{\nu}} \; .
\end{displaymath}
We have computed $\mathrm{Im} (\kappa_0)$ on different lattices and
extracted $\nu$ from a double logarithmic plot. Our results are
consistent with those obtained by the other methods but the
statistical errors again turned out to be substantially larger than
for the Binder method.

\section{Magnetic transition at vanishing Yukawa coupling}

\subsection{The $\phi_3^4$ model}

In the limit $y=0$ the action (\ref{Gitterwirk}) describes free
massless fermions and $O(2)$ invariant $\phi_3^4$ model with quartic
selfcoupling.  Besides our interest in the features of the $\phi_3^4$
model as a limit of the Yukawa theory, here we have developed and
tested the methods we wanted to apply to the more sophisticated and
expensive fermionic model.  The existence of a nontrivial fixed point
and a finite nonvanishing value of the renormalized quartic
selfcoupling $\lambda_R$ in the continuum limit make this model by
itself very interesting from a field theoretic point of view, too.

\begin{figure}[htbp]
  \begin{center}
    \leavevmode
    \psfig{file=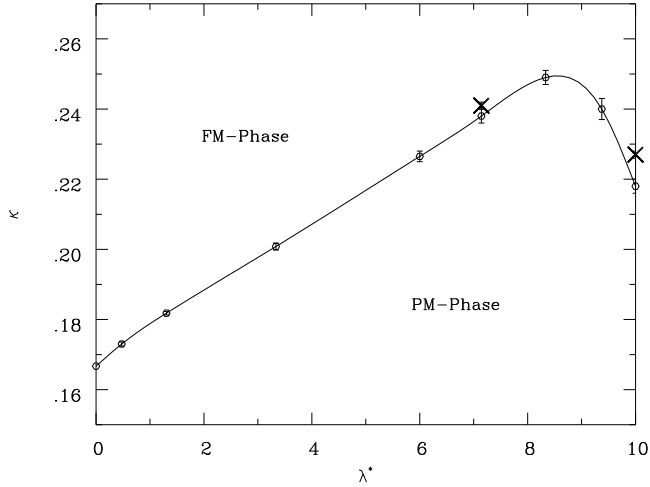,angle=90,height=8cm}
  \end{center}
  \caption{Phase diagramm of the $\phi_3^4$ model. The
    $\lambda$-axis has been rescaled:
    $\lambda^*=50\lambda/(1+5\lambda)$.
    The circles show the maxima of the susceptibility on a $6^3$
    lattice. The crosses show the positions of $\kappa_c$ determined
    with higher precision by means of finite size scaling methods.}
  \label{FIG:BOS:PD}
\end{figure}

The phase diagram in the $\kappa$-$\lambda$ plane, computed mainly on $6^3$
lattices (see the $y=0$ entries in the table \ref{Peaks}), is displayed in
fig.~\ref{FIG:BOS:PD}. The spectrum in the
PM phase below the
second order phase transition line contains two degenerate massive scalar
particles. In the FM phase ($\kappa > \kappa_c$) the $O(2)$
symmetry is spontaneously broken and the lightest particles in the spectrum are
a massless Goldstone boson and a massive $\sigma$ boson.

The renormalization group properties of the $\phi_3^4$ model have been
investigated e.g. in \cite{BrLe73} and are indicated in
figs.~\ref{FIG:RGFL} and \ref{FIG:RGFL2} on the $y=0$ face of the
phase diagram at $\kp>0$. The model is superrenormalizable in weak
coupling perturbation theory and its physics at infinitesimal scalar
selfcoupling is dominated by the Gaussian fixed point (Gfp) at $\lambda=0$.
                                %
At nonvanishing coupling $\lambda>0$ the critical line
$\kappa_c(\lambda)$ is dominated by the IR-stable nontrivial WFfp.
The investigations by means of $\varepsilon$ expansion or $1/N$
expansion of the $O(N)$ symmetric $\phi_3^4$ suggest that the
interaction term becomes irrelevant, and the only relevant term
remains the kinetic one.  This means that at $\lm>0$ only one parameter
has to be tuned $\kp\ra\kp_c(\lm)$ in order to reach a continuum
limit governed by the WFfp. Thus the same scaling behavior should be
found when the critical line is approached at arbitrary $\lm > 0$.

\subsection{Results at ${\mathbf \lambda=\infty}$ and ${\mathbf
    \lambda=0.5}$}

We have chosen $\lm = \infty$ and $\lm = 0.5$ and determined the renormalized
coupling as well as some critical indices in runs in the $\kp$ direction.
A Monte Carlo determination of the renormalized quartic coupling $\lm_R$ has
been done e.g. in \cite{FrSm82} for the $Z(2)$ symmetric $\phi_3^4$ model.
To our knowledge no analogous measurement exists for the $O(2)$
symmetric model. Following e.g. \cite{FrSm82} we define $\lm_R$ in
the symmetric phase as
\begin{equation}
  \lambda_R=(Lam_R)^3 U_L~.
  \label{BOS:RC}
\end{equation}
Here $am_R$ is the mass of the $\sg$-boson extracted from the scalar
propagator. To extrapolate to the continuum limit we varied the
lattice size from $L=6$ to $12$ while keeping $Lam_R$ fixed to $4$.

At $\lambda=\infty$ the renormalized scalar selfcoupling increases
very slowly with the lattice size $L$. The linear extrapolation in
$1/L$ to $L=\infty$ suggests a value of $\lambda_R=26\pm4$.
At $\lambda=0.5$ an extrapolation to $L=\infty$ is less precise,
suggesting $\lm_R=20-30$. The agreement supports the expectation, that
the model is dominated by the WFfp on the whole critical line
$\lm>0$. These results for $\lm_R$ are also consistent with the
expected theoretical value \cite{Zi89}.

\begin{figure}[htbp]
  \begin{center}
    \leavevmode
    \psfig{file=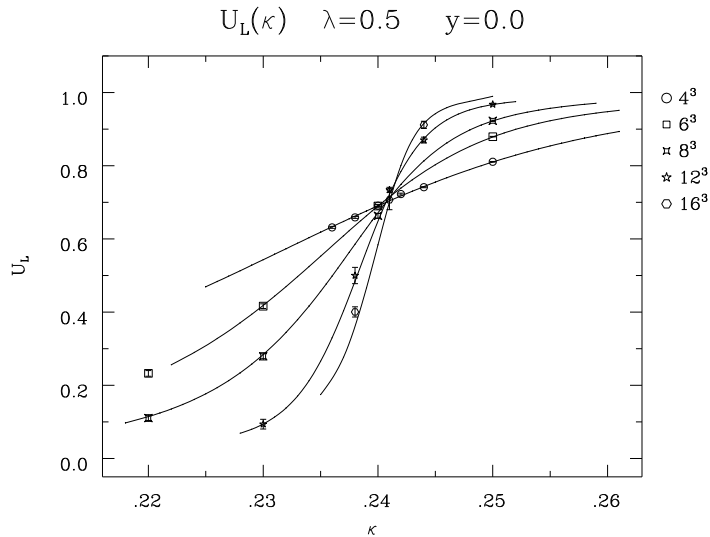,angle=90,height=8cm}
  \end{center}
  \caption{The intersection point of the Binder cumulants in the
    $\phi^4_3$ model on several lattices  for $\lambda=0.5$ at
    $\kappa_c=0.241(1)$. The lines were obtained by reweighting, the
    symbols are the measured points.}
  \label{FIG:BOS:KC}
\end{figure}

\begin{figure}[htbp]
  \begin{center}
    \leavevmode
    \psfig{file=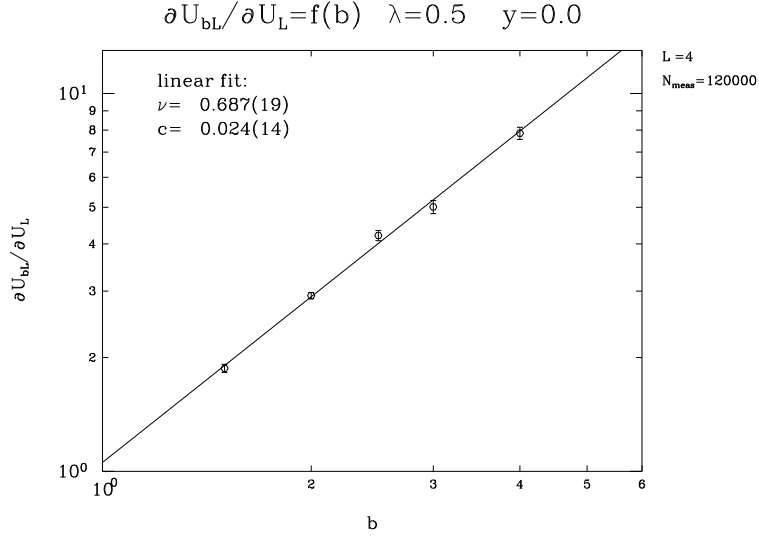,angle=90,height=8cm}
  \end{center}
  \caption{Linear fit to $\ln(\partial U_{bL}/\partial U_L)$ used,
    according to eq.~\ref{NuBinderA}, for the determination of the
    critical exponent $\nu$ determined at $\lambda=0.5$.}
  \label{FIG:BOS:NU}
\end{figure}

The most sensitive test for the appurtenance to the same universality
class is the comparison of critical exponents. Using the Binder method
described in subsec.~3.1 we have determined the critical exponents
$\nu$, $\beta/\nu$ and $\gamma/\nu$.  The method works very well at
both $\lm$ values in the whole range of lattice sizes used,
$4^3-16^3$. To illustrate this we show in fig.~\ref{FIG:BOS:KC} the
determination of $\kp_c$ at $\lm=0.5$. In fig.~\ref{FIG:BOS:NU} the
data for $\partial U_{bL}/\partial U_L$, used for the determination of
$\nu$ at the same $\lm$ value, and the linear fit, are displayed.

\begin{table}[htbp]
  \begin{center}
    \leavevmode
    \begin{tabular}{|c||c|c|c|c|c|c|c|}
      \hline
      $\lambda$ & $\kappa_c$ & $\nu$ & $\beta/\nu$ & $\gamma/\nu$ &
      $(\alpha)$ & $(\delta)$ & $(\eta)$ \\
      \hline
      $\infty$ & 0.2275(10) & 0.673(19) & 0.51(3) & 2.03(6) &
      -0.02(6) & 5.2(4) & 0.02(6) \\
      0.5 & 0.241(1) & 0.687(19) & 0.56(5) & 1.91(6) & -0.06(6) &
      4.5(3) & 0.12(10) \\
      \hline
    \end{tabular}
  \end{center}
  \caption{Critical exponents in the $\phi_3^4$ model at $\lambda=0.5$
    and $\lambda=\infty$. The exponents enclosed in brackets were
    calculated by using hyperscaling relations.}
  \label{TAB:BOS:CE}
\end{table}

We were able to determine
$\nu$ to a precision of about $3\%$, $\beta/\nu$ to about $9\%$ and
$\gamma/\nu$ to $3\%$. The results are summarized in table
\ref{TAB:BOS:CE}.
They are consistent with the expectation that the
two points $\lambda=\infty$ and $\lambda=0.5$ belong to the same
universality class.
The value of $\beta/\nu$ is consistent with the one calculated with
the hyperscaling relation $\beta/\nu=(d-\gamma/\nu)/2$. As
hyperscaling seems to be fulfilled, we determined the exponents
$\alpha$, $\delta$ and $\eta$ from the relations
\begin{displaymath}
  \alpha=2-\nu d,\quad \delta=\frac{d+\gamma/\nu}{d-\gamma/\nu},\quad
  \eta=2-d+2\frac{\beta}{\nu}~.
\end{displaymath}

\section{Chiral transition at vanishing scalar selfcoupling}

\subsection{The GN$_3$ model}

At $\lambda = \kappa = 0$ the scalar field $\phi$ plays in the action
(\ref{Gitterwirk}) the role of an auxiliary field. It can be
integrated out thus obtaining a purely fermionic GN$_3$ model with U(1)
chiral symmetry,
\begin{equation}
 S=S_F-\frac{y^2}{4} \left[
   \left( \frac{1}{8}\sum\limits_{b\in h.c.} \chb_{x+b}\chi_{x+b}
   \right)^2
  -\left( \frac{1}{8}\sum\limits_{b\in h.c.}
    \epsilon_{x+b}\chb_{x+b}\chi_{x+b} \right)^2
  \right].
 \label{GN:ACTION}
\end{equation}
In 3D this model is perturbatively non-renormalizable. However, it has
been shown in \cite{RoWa89,RoWa91} that the GN$_3$ model is renormalizable
in the $1/N_F$ expansion. The $\beta$-function has been calculated to
$O(1/N_F)$ in \cite{HaKo93,HeKu92} and to $O(1/N_F^2)$ in \cite{Gr94}. The
$1/N_F$ expansion reveals a nontrivial UV-stable fixed point where
dynamical chiral symmetry breaking and fermion mass generation occur. The
phase transition is of $2^{\rm nd}$ order and the order parameter is the
chiral condensate $\langle\cbc\rangle$.

In ref.~\cite{Gr94} one can find the critical exponent $\nu$ to
$O(1/N_F^2)$. In our case ($N_F=4$)
\begin{equation}\label{mooo}
  \nu=1+\frac{16}{3\pi^2 N_F}-\frac{8(376+27\pi^2)}{27\pi^4 N_F^2}
  +O(1/N_F^3) \simeq 1+0.135-0.122 \simeq 1.0(1)~.
\end{equation}
The $O(1/N_F)$ term is identical with the results in
\cite{HeKu92,HaKo93}. The $O(1/N_F^2)$ term is of the same order of
magnitude, which suggests a rather large error on the value of $\nu$
in (\ref{mooo}).

In the symmetric phase ($y<y_c$) fermions are massless. This region
is dominated by the trivial Gaussian fixed point at $y=0$.

By adding the kinetic scalar term to the bare GN$_3$ action the scalar
field $\phi$ turns from an auxiliary field to a dynamical one. This
restricted Yukawa model with $\lm=0$, sometimes considered as a
sufficient representation of the $Y_3$ model (e.g. in \cite{GaKo92}),
is a natural extension of the parameter space of the GN$_3$ model. We
know that such a Yukawa model with vanishing scalar selfcoupling and
$Z(2)$ symmetry is renormalizable in $1/N_F$ expansion.  As shown in
\cite{Zi91,KoRo91}, this model has a nontrivial IR-stable fixed point
where the kinetic term of the scalar field becomes irrelevant and the
4-fermion interaction term relevant. This fixed point is identical
with the critical GN$_3$ model. The IR-stable fixed point of this
restricted Yukawa model corresponds to the UV-stable fixed point of
the GN$_3$ model \cite{Zi91,KoRo91,GaKo92}. This can be understood
from the renormalization group flow in a larger parameter space, in
the full Y$_3$ model (\ref{Gitterwirk}) (see figure \ref{FIG:RGFL2}).
The flow is suggested by the $\beta$-functions obtained in the
$\epsilon$-expansion \cite{Zi91}.  The RG-flow restricted to the
GN$_3$-line $\kp=\lm=0$ is consistent with the UV-stability of the
nontrivial GN$_3$ fixed point.

\subsection{Numerical results}

First we comment on the spectrum calculations.
The fermion mass $am_F$ has been measured by fitting the momentum space
fermion propagator, measured usually at four lattice momenta, to a
free fermion ansatz. In the broken phase $am_F$ agrees very
well with the tree level relation $am_F=y\lag\phi\rag =
\frac{y^2}{2}\lag\cbc\rag$.

For the measurement of the masses of the $\sigma$-boson $am_\sigma$ and
the Goldstone boson $am_\pi$ we had to use an ansatz for the momentum
space propagators from the one-loop
renormalized perturbation theory \cite{Frick}. In this case the
previously fitted fermion mass is used to calculate the fermionic
selfenergy which contributes to the renormalized bosonic propagators.
This method delivers the renormalized Yukawa coupling and describes
very well the form of the bosonic propagators which differ very much
from the free ones. As expected, in the FM phase $am_\pi$ is very
small and $am_\sg$ increases with the distance from the critical
point. In the PM phase both masses grow with the distance from the
critical point and become degenerate.

The scaling behavior has been investigated in two directions: In the
GN$_3$ case ($\kp=0$) we varied $y$ and determined the critical Yukawa
coupling $y_c=1.091(5)$ from the intersection point of the Binder
cumulants on several lattices. From the finite size scaling behavior
of the Binder cumulant at this value we determined the exponent $\nu$.
Similarly the behavior of magnetization and susceptibility allowed us
to determine $\beta/\nu$ and $\gamma/\nu$, respectively. The obtained
results are collected in table \ref{TAB:GN:CE}.

\begin{table}[htbp]
  \begin{center}
    \leavevmode
    \begin{tabular}{|c|c||c|c|c|c|}
      \hline
      $\kappa_c$ & $y_c$ & $\nu$ & $\beta/\nu$ & $\gamma/\nu$ & note\\
      \hline
      0 & 1.091(5) & 1.02(8) & 0.89(10) & 1.19(13) & run in $y$ (GN)\\
      0.000(2) & 1.09 & 1.05(12) & 0.90(4) & 1.15(4) & run in $\kappa$ \\
      \hline
    \end{tabular}
  \end{center}
  \caption{The critical exponents $\nu$, $\beta/\nu$, $\gamma/\nu$ in
    the GN$_3$ model and in the Yukawa model at vanishing $\lambda$.}
  \label{TAB:GN:CE}
\end{table}

By using the measured $\gamma/\nu$ value one obtains from the
hyperscaling relations $\beta/\nu=0.905(65)$. This is in good
agreement with the measured value and supports the hyperscaling
hypothesis.

\begin{figure}[htbp]
  \begin{center}
    \leavevmode
    \psfig{file=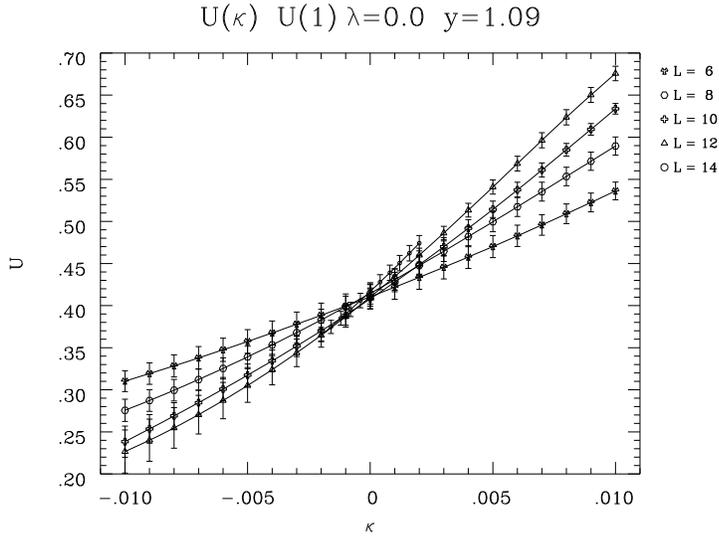,angle=90,height=8cm}
  \end{center}
  \caption{Determination of $\kappa_c$ in the Yukawa model at
    $\lambda=0$ and $y=1.09$.  The intersection point of the BCL
    cumulants measured on different lattice sizes gives
    $\kappa_c=0.000(2)$ which is in perfect agreement with the GN$_3$
    critical point.}
  \label{FIG:GN:KC}
\end{figure}

\begin{figure}[htbp]
  \begin{center}
    \leavevmode
    \psfig{file=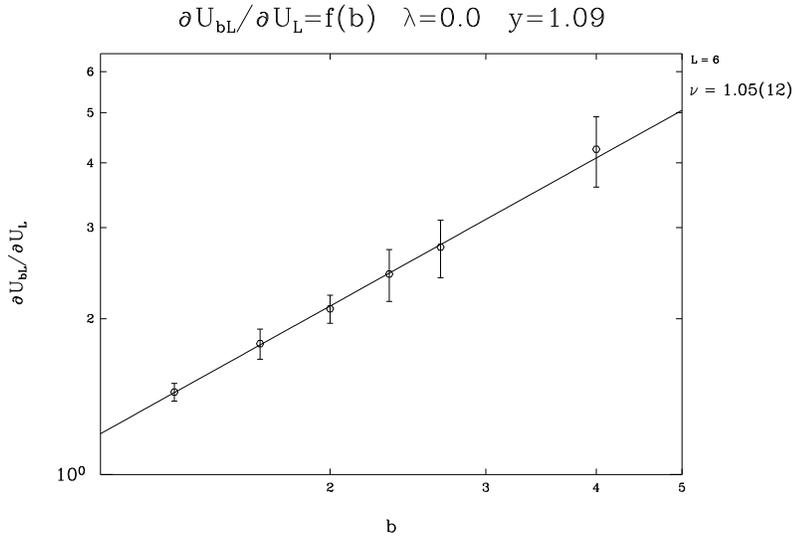,angle=90,height=8cm}
  \end{center}
  \caption{Determination of the critical exponent $\nu$ in the Yukawa
    model at $\lambda=0$, $y=1.09$ and $\kappa=0$.}
  \label{FIG:GN:NU}
\end{figure}

As a test of our methods and of the equivalence between the fixed
points of the GN$_3$ and the Yukawa model with vanishing $\lambda$ we
measured the critical exponents in the latter model by approaching the
critical point of the GN$_3$ model along the $\kappa$ direction.
Fig.~\ref{FIG:GN:KC} demonstrates that the critical point obtained by
the Binder method in this direction is identical with the GN$_3$ one.
The BCL cumulants intersect at $\kappa_c=0.000(2)$ and $y_c=1.09$. As
shown in fig.~\ref{FIG:GN:NU}, the values of $\beta/\nu$ and $\gamma/\nu$
are perfectly consistent with those obtained in the GN$_3$ run (table
\ref{TAB:GN:CE}).

We conclude that the Binder finite size scaling method is applicable
and gives consistent results in the Yukawa model at $\lm=0$ for a
broad range of lattice sizes.  The values of the critical exponents in
the chiral GN$_3$ model are the same as in the Yukawa model with
$\lm=0$. This confirms that the fixed points of
these two models are the same. The exponents are consistent with the
$1/N_F$ predicted values ($\nu\approx 1$) and significantly different
from the exponents associated with the WFfp. This allows us to
investigate the crossover effects between these universality classes
numerically.

\section{Gross-Neveu-like behaviour for strong couplings}

The $1/N$ expansion predicts \cite{Zi91,KoRo91,GaKo92} that the Y$_3$
model and the GN$_3$ models are equivalent at least for weak scalar
selfcoupling $\lambda$. In order to test this hypothesis also for
strong scalar couplings we have investigated the Y$_3$ model with
$\lambda = \infty$ at strong bare Yukawa coupling $y=1.1$. This choice
leads to $\kp_c\simeq 0$.

\begin{figure}[htb]
  \begin{center}
    \leavevmode
    \psfig{file=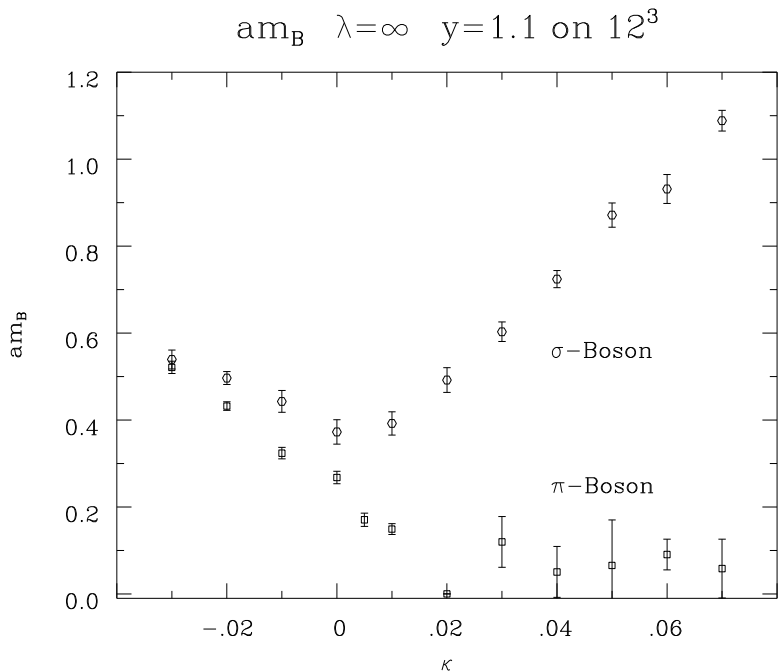,height=10cm,width=14cm,angle=90}
  \end{center}
  \vspace{-1.2cm}
  \caption{\protect\label{BosMassy1.1} The masses of the $\sg$- and
    $\pi$-bosons as functions of $\kappa$. As expected, in the broken phase the
    $\sigma$ boson is massive and the $\pi$ boson massless.}
\end{figure}

The spectrum is similar to that of the GN$_3$ model.  We observe the
generation of the fermion mass $am_F$ which is related to a nonzero
chiral condensate $\lag\cbc\rag$. Even for $\lambda = \infty$, where
the $1/N_F$-expansion is not applicable, we find that the prediction
$am_F \approx y \langle \phi \rangle$ is fulfilled with good
precision.  Fig.~\ref{BosMassy1.1} displays the dependence of the
masses of both bosons on the hopping parameter $\kappa$ at $y=1.1$ and
$\lambda = \infty$.  In accordance with the Goldstone theorem one
massive $\sg$ boson and one massless $\pi$ boson appear in the FM
phase.  The qualitative $\kp$-dependence of both masses in the
vicinity of the critical point is the same as in the Y$_3$ model at
$\lambda=0$. This is a first numerical hint for the physical
equivalence of both cases.

In order to determine the universality class of the Y$_3$ model at
$\lm=\infty$ and strong Yukawa coupling we have again determined the
critical exponents $\nu$, $\beta/\nu$ and $\gamma/\nu$. We have applied the
Binder method at $y=1.1$ approaching the critical sheet in the $\kp$
direction.
The critical value $\kappa_{\mathrm{c}}=0.007(2)$ is given by the
common intersection point of the cumulants $U_L$ on different
lattices sizes $L$.

\begin{figure}
  \begin{center}
    \leavevmode
    \psfig{file=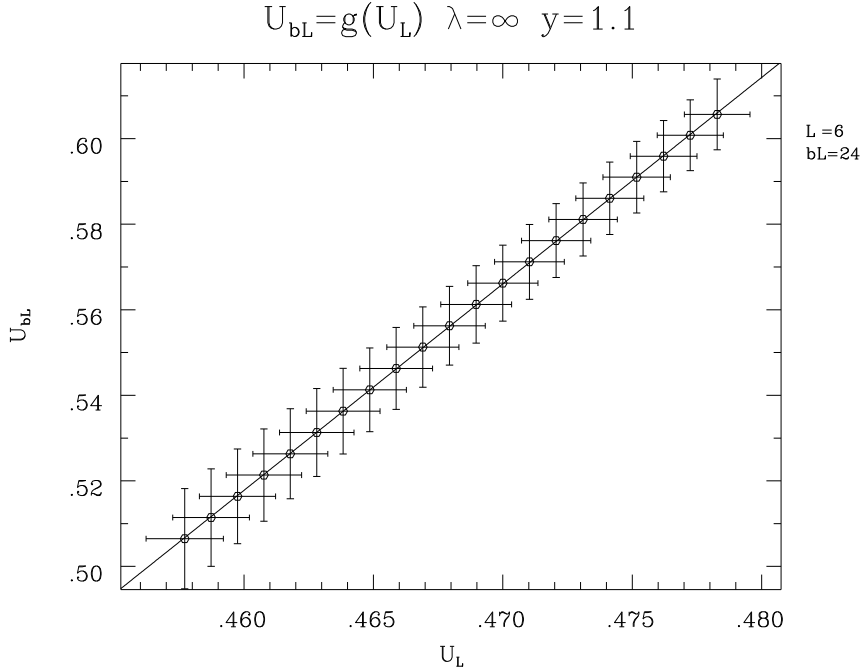,height=10cm,width=14cm,angle=90}
  \end{center}
  \vspace{-1.2cm}
  \caption{An example ($L=6$, $b=4$) of the linear dependence of
    $U_{bL}$ on $U_L$ at $\lm=\infty$, $y=1.1$ near the critical point.}
  \label{cumulfunc}
\end{figure}

For this value of $\kappa_{\mathrm{c}}$ we have computed the
derivatives $\partial U_{bL} / \partial U_L$ with $L=6$ and $bL$
ranging from 8 to 24. Figure~\ref{cumulfunc} shows as an example
$U_{bL}$ as a function of $U_L$ for $b=4$. Near the critical point such
functions are linear with good precision and the derivatives are thus
easily determined.

\begin{figure}
  \begin{center}
    \leavevmode
    \psfig{figure=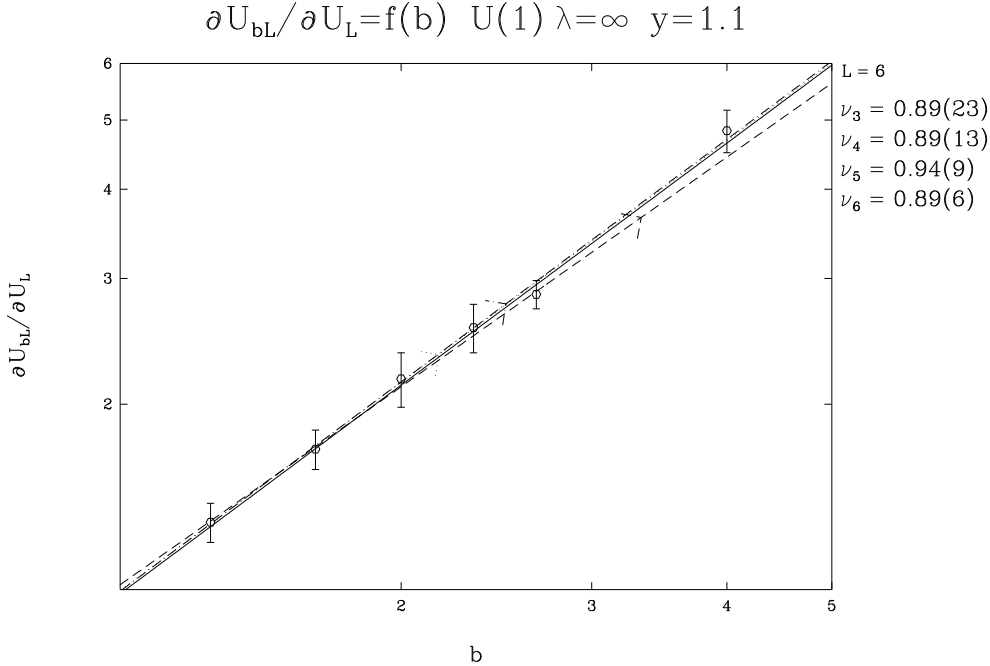,height=10cm,width=14cm,angle=90}
  \end{center}
  \caption{Log-log plot of the derivative
    $\partial U_{bL} /\partial U_L$ at $\kappa_{\mathrm{c}}$ at
    $\lambda=\infty$, $y=1.1$. We show the linear fits to the first
    three,$\dots$, six points.
    The results of these fits are consistent values
    $\nu_3,\ldots,\nu_6$ of the exponent $\nu$.}
  \protect\label{GNlikeNu}
\end{figure}

Using equation (\ref{NuBinderA}) we have obtained the critical exponent
$\nu$,
\begin{equation}
\nu = 0.89(6) \; .
\end{equation}
This value is a little bit smaller than the one obtained at $\lm=0$,
but both values are consistent within statistical errors. Figure
\ref{GNlikeNu} shows the corresponding plot. We have also made various
fits with different subsets of data points.  The results are nearly
unaffected if we leave out one or more data points in the fit. This
shows that also for $\lambda = \infty$ and $y=1.1$ corrections to
scaling are quite small and the Binder method works in a broad range
of lattice sizes.

We have further determined the ratios $\beta/\nu$ and $\gamma/\nu$,
\begin{eqnarray}
\frac{\beta}{\nu} &=& 0.80(8), \nonumber \\
\frac{\gamma}{\nu} &=& 1.30(7) \; .
\end{eqnarray}
Within statistical errors these exponents are consistent with our
results in the GN$_3$ model, too. They fulfill the corresponding
hyperscaling relation with good precision.

These numerical results lead us to the conclusion that the Gross-Neveu
universality class extends over the whole range from $\lambda=0$ to
$\lambda=\infty$ provided the bare Yukawa coupling $y$ is strong
enough, $y\simeq 1$. This confirms the conjecture that the GN$_3$
model and the $\mathrm{Y}_3$ model are equivalent field theories even
for $\lambda=\infty$.

\section{Interplay of magnetic and chiral universality classes}

Both in the pure scalar $\phi^4_3$ theory at $y = 0$ and in the GN$_3$
model at $\lm = 0$ the Binder method works in an exemplary way.  Also
in the Y$_3$ model at $\lm = \infty$, $y = 1.1$ it provides
satisfactory results. This is presumably due to the dominance of only
one of the fixed points in these cases. They seem to be ``pure''
cases, without any interplay of universality classes.  Now we
describe what happens in the Y$_3$ model when at $\lm = \infty$ the
Yukawa coupling $y$ is decreased, and the XY$_3$ model is approached.
We made extensive simulations at $y = 0.6$ and $y = 0.3$, approaching
the critical sheet in the $\kp$ direction.

\subsection{${\mathbf \lm = \infty}$, ${\mathbf y = 0.6}$}

\begin{figure}[htbp]
  \begin{center}
    \leavevmode
    \psfig{file=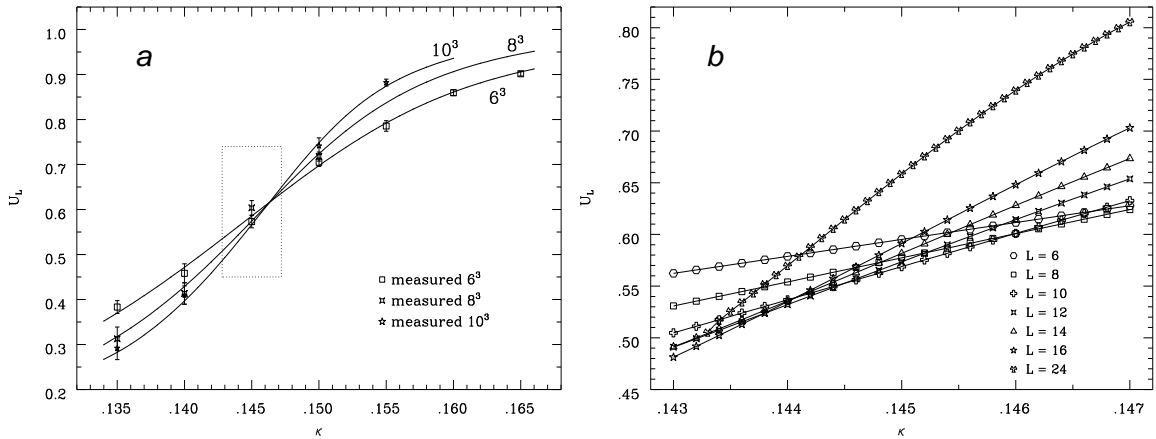,width=\linewidth}
  \end{center}
  \caption{Determination of $\kappa_c$ at $\lambda=\infty$ and
    $y=0.6$. On small lattices (a) the apparent critical $\kp$ is
    0.1463, but including also data on larger lattices and zooming
    into the rectangular region (b) suggests that the critical point
    is in the interval $\kp=0.1435-0.1445$ of the
    intersection points of $U_L$ for $L\geq 10$.}
  \label{FIG:SC:KC}
\end{figure}

For small lattice sizes, $L = 6, 8, 10$, the cumulants consistently
cross in the interval $\kp = 0.1460 - 0.1466$ (fig.
\ref{FIG:SC:KC}a).  Making the finite size analysis at $\kp = 0.1463$
we obtain $\nu = 0.75(9)$, a value quite close to that of the XY$_3$
model.

However, when only large lattices $L = 10, 12, 14, 16, 24$ are considered,
the crossing point is found in the interval $\kp = 0.1430 - 0.1445$.
The situation is shown on a fine $\kp$ scale in fig. \ref{FIG:SC:KC}b.
For these lattices at $\kp = 0.144$ we find $\nu = 0.99(23)$,
a value consistent with the GN$_3$ model, but with a large error.

\begin{figure}[htbp]
  \begin{center}
    \leavevmode
    \psfig{file=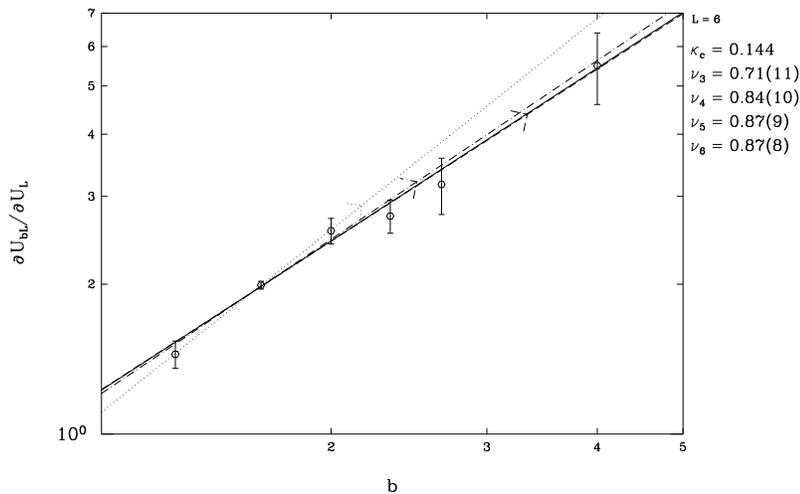,angle=90,height=8cm}
  \end{center}
  \caption{Similar to fig.~\protect\ref{GNlikeNu}, but now at
    $\lm=\infty$, $y=0.6$. The $\nu$-values increase systematically
    when data on larger and larger lattices is included in the linear fit.}
  \label{FIG:5_9}
\end{figure}

We have made an analysis at $\kp_c = 0.144$ including data on all
lattices and choosing the basis $L = 6$.  The $\nu$ values have been
determined for different groups of data, for the first 3, 4, 5 and 6
points.  As shown in fig. \ref{FIG:5_9}, when data on larger and
larger lattices is included, $\nu$ increases systematically from
0.71(11) for $b = 8/6$, 10/6 and 12/6 only, to 0.87(8) when all data is
included.  This is probably not a good way of analysis in such a
complex situation and the previous one made only on large lattices
seems to be more reliable.  We have made it in order to illustrate the
systematic increase of the apparent $\nu$ with lattice size.

We interpret the above results as a hint that
for sufficiently large lattices, $L \ge 10$,
the $\chi$fp universality class finally shows up.
It might be tempting to conjecture that the low value of $\nu$,
obtained rather consistently for $L \le 10$,
is a signal for the nearby WFfp class.
But, as the results at $y = 0.3$ indicate, this is questionable.

\begin{figure}[htbp]
  \begin{center}
    \leavevmode
    \psfig{file=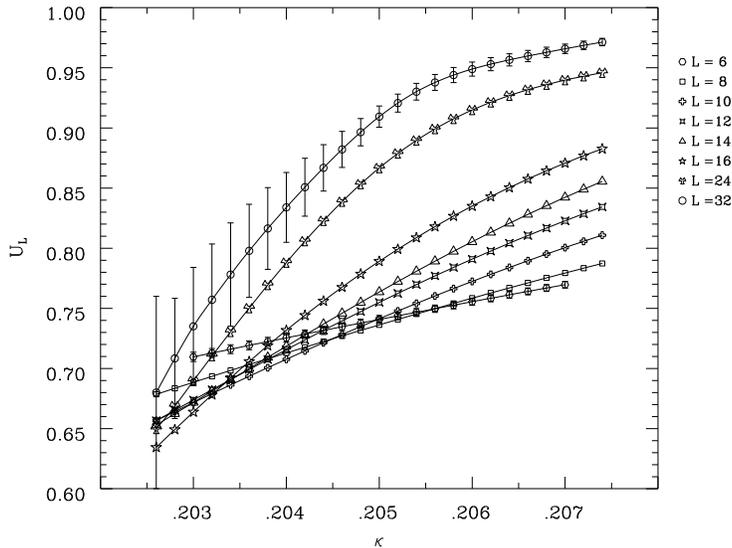,angle=90,height=8cm}
  \end{center}
  \caption{An attempt to determine $\kappa_c$ at $\lambda=\infty$,
    $y=0.3$ failed. $U_L$ do not intersect in a single point even if
    only large lattices are considered.}
  \label{FIG:SC:KC03}
\end{figure}

\begin{figure}[htbp]
  \begin{center}
    \leavevmode
    \psfig{file=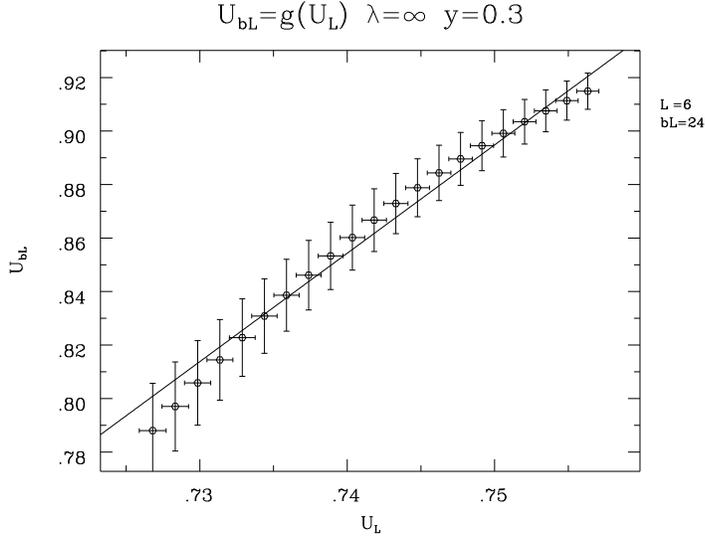,angle=90,height=8cm}
  \end{center}
  \caption{Example ($L=6$, $b=4$) of nonlinear dependence of $U_{bL}$
    on $U_L$ at $\lm=\infty$, $y=0.3$ near the critical point.}
  \label{FIG:SC:DU}
\end{figure}

\subsection{${\mathbf \lm = \infty}$, ${\mathbf y = 0.3}$}

As seen in fig.~\ref{FIG:SC:KC03}, the cumulants obtained on lattices
up to $L = 32$ show no tendency to cross at some unique point, even if
smaller lattices are discarded.  Also the dependence of $U_{\rm bL}$
on $U_{\rm L}$, shown in fig.~\ref{FIG:SC:DU}, is not linear,
differing e.g. from $\lm=\infty, y=1.1$, seen in fig.~\ref{cumulfunc}.
A determination of $\nu$ under these circumstances makes little sense,
and one can only speculate that if lattices could be made still
substantially larger, a simpler finite size behavior with the $\chi$fp
exponents might be found.

Remarkable is also the fact that the finite size behavior
did not improve on small lattices.
As in the $y = 0.6$ case, the cumulants on $L \le 10$ lattices
cross in a narrow interval $\kp_c = 0.2045 - 0.2055$.%
\footnote{The corresponding value of $\nu$ is $\nu \simeq 0.76$
with errors difficult to estimate because of systematic
uncertainties caused e.g. by a nonlinearity of
the dependence of $U_{\rm bL}$ on $U_{\rm L}$.}
But including the $L = 12$ data spoils the consistency completely.
Thus halving the distance from the XY$_3$ model with respect to $y = 0.6$
did not increase the consistency of the finite size behaviour
for smaller lattice sizes.
This prevents us from interpreting the low values of $\nu$
obtained on smaller lattices as a signal
for the WFfp universality class.

Attempts to incorporate some corrections to the leading finite size
behaviour, as suggested in ref. \cite{BeGo88b}, are in our case not
very helpful because simulations with dynamical fermions cannot yet
produce data with the precision needed to deal with additional
parameters.  Thus, we conclude that at $\lm = \infty$ and $y = 0.3$
the finite size behaviour is not under control.  Unfortunately, it
would not have been easy to notice that without having data in a large
range of lattice sizes.

\section{Summary and Conclusions}

We have studied the finite size behavior of the Y$_3$ model with U(1) chiral
symmetry along the 2 dimensional sheet of chiral phase transitions at various
values of the Yukawa coupling $y$ and of the scalar selfcoupling $\lm$.  The
aim was to investigate the influence and the interplay of the two nongaussian
fixed points of the model for various values of the couplings.

In the $y=0$ limit case, i.e. in the $\phi^4_3$ model, the critical
exponents associated with the Wilson-Fisher fixed point (WFfp) are
clearly observed both at $\lm=\infty$ and $\lm = 0.5$. The Binder
method of finite size scaling analysis is applicable already on small
lattices. Also the renormalized coupling values agree and are
consistent with the theoretical prediction. The WFfp thus provides a
rather complete description of the model at least for $\lm \ge 0.5$.

For $y>0$ we find that the chiral fixed point ($\chi$fp) determines
the finite size scaling in the vicinity of the chiral phase transition
sheet as long as the Yukawa coupling is strong enough, $y\simeq 1$.
The independence on the value of the quartic coupling $\lm$ confirms
the expectation that the $Y_3$ model and its special case, the GN$_3$
model, belong to the same universality class of the $\chi$fp. Also the
fermion and boson masses at $\lm=0$ and $\lm=\infty$ are very similar.
For $y\simeq 1$ the Binder method of finite size scaling analysis
works consistently in a broad range of lattice sizes, in analogy to
the pure $\phi_3^4$ theory. No substantial difference in the finite
size behavior has been found between $\lm=0$ and $\lm=\infty$. This
implies that as long as $y$ is large enough, the $\lm\phi^4$ term does
not influence the finite size behavior of the Y$_3$ model even on
small lattices and the model is rather completely described by the
$\ch$fp.

When, at $\lm=\infty$, $y$ is decreased to $y = 0.6$, the finite size
behavior cannot be analysed any more by the Binder method in the whole
range of the lattice sizes we used. The behavior on small ($L \le 10$)
and large ($L \ge 10$) lattices is different. On the larger lattices
the $\ch$fp seems still to dominate. On the smaller lattices the
behavior looks consistent with the WFfp.  But this does not
necessarily mean that the WFfp already starts to show up: when a
further step towards the $\phi^4_3$ limit case is made, at
$\lm=\infty$ and $y=0.3$, the finite size behavior does not show
increased resemblance to that fixed point. If applied in a narrow
interval of lattice sizes, the Binder method might seem to be
applicable but the results are misleading.

A numerical verification of the expectation that the Y$_3$ model is
equivalent to the GN$_3$ model is thus very difficult for $y \le 0.6$.
Our tentative conclusion is that the observed deviation from the
simple finite size scaling signals an interplay of both universality
classes, i.e. a crossover between them. This warns us that in the
situation of intertwining phenomena the finite size behavior may be
very complex.  As we learned in the $\lm=\infty, y=0.6$ case, this
fact is not easily detectable in a small range of lattice sizes,
however.

\section*{Acknowledgements}

We thank W. Franzki, M. G\"ockeler, S. Hands, P. Hasenfratz and M.M.
Tsypin for discussions, K. Binder for comments and some informations,
P. Lacock for the reweighting program and H.A. Kastrup for support.
The simulations have been performed on the Cray YMP of the HLRZ
J\"ulich (small lattices) and the Quadrics QH2 of the DFG in Bielefeld
(large lattices).

\end{document}